\documentclass[aps,prl,superscriptaddress,twocolumn]{revtex4}

\usepackage{amsmath,amssymb}
\usepackage{color}
\usepackage{hyperref}
\usepackage{graphicx}
\usepackage{subfig}

\newcommand{\bea}{\begin{eqnarray}}
\newcommand{\eea}{\end{eqnarray}}
\newcommand{\be}{\begin{equation}}
\newcommand{\ee}{\end{equation}}

\def\b{\beta}       
     
\def\d{\delta}  \def\D{\Delta}  
        
\def\f{\phi}

\def\m{\mu}

\def\r{\rho}

\def \Tr {{\textrm{Tr}}}

\def \be {\begin{equation}}
\def \ee {\end{equation}}
\def \ba {\begin{array}}
\def \ea {\end{array}}
\def \bea{\begin{eqnarray}}
\def \eea{\end{eqnarray}}

\def \b {\beta}

\def \d {\delta}
\def \D {\Delta}

\def \m {\mu}

\def \r {\rho}

\def \f {\frac}

\begin{document}
%\begin{CJK*}{GBK}{song}

\title{Universal relation between thermal entropy and entanglement entropy in CFT}

\author{Bin Chen}
\email[Email: ]{bchen01@pku.edu.cn}
\affiliation
{Department of Physics, and State Key Laboratory of Nuclear Physics and Technology, Peking University, Beijing 100871, P.R. China}
\affiliation{Collaborative Innovation Center of Quantum Matter, Beijing 100871, P.~R.~China}
\affiliation{Center for High Energy Physics, Peking University, Beijing 100871, P.R. China}
\author{Jie-qiang Wu}
\email[Email: ]{jieqiangwu@pku.edu.cn}
\affiliation
{Department of Physics, and State Key Laboratory of Nuclear Physics and Technology, Peking University, Beijing 100871, P.R. China}

\begin{abstract}

Inspired by the holographic computation of large interval entanglement entropy of two-dimensional conformal field theory at high temperature, it was proposed that the thermal entropy is related to the entanglement entropy as $S_{th}=\displaystyle{\lim_{l \to 0}(S_{EE}(R-l)-S_{EE}(l))}$. In this letter, we prove this relation for 2D CFT with a discrete spectrum in two different ways. Moreover we discuss this relation  for a 2D noncompact free scalar, which is a gapless CFT with continuous spectrum. We show that it could be recovered, after appropriately regularizing the theory.
%{PACS}: 04.70.Dy, 11.25.Hf

\end{abstract}

\maketitle

\section{Introduction}

The entanglement is a uniquely  quantum mechanical property and plays an important role in understanding the quantum many body  systems. To measure the entanglement in a bipartite system, one may define the entanglement entropy as the von Neumann entropy of the reduced density matrix of subsystem $A$\cite{nielsen2010quantum},
\be
S_A=-\Tr_A \r_A\log \r_A.
\ee
Here the reduced density matrix is obtained by smearing over the degrees of freedom of subsystem $B$ complementary to $A$. If the system is in a pure state, one has
$S_A=S_B$.
However, if the system is at finite temperature,  due to the thermal effect we get $S_A \neq S_B$.
One may define the quantity $\d S =S_A-S_B$ to measure the deviation from purity, which is bounded by the Araki-Lieb inequality\cite{Araki:1970ba}
\be
|\d S|=|S_A-S_B|\leq S_{A\cup B}.
\ee
Obviously, if $A$ is the whole system, then the above inequality saturates and the entanglement entropy reproduces exactly the thermal entropy of the system. The study of entanglement at finite temperature sheds light on the interplay between the quantum nature of the system and its thermodynamics.  % (it needs some modification for continuous system).

Among various studies on the entanglement entropy  in  many-body systems(see \cite{Amico:2007ag} for a nice review), the one in quantum field theory is of particular interest. As the quantum field encodes an infinite number of degrees of freedom, its vacuum is highly entangled. In this case, the entanglement entropy is called geometric entropy as its leading contribution satisfies an area law\cite{Eisert:2008ur}.

Quite recently the entanglement entropy opened a new window to study AdS/CFT correspondence. In \cite{Ryu:2006bv,Ryu:2006ef}, it was proposed that the entanglement entropy of submanifold $A$ in a conformal field theory(CFT) could be holographically given by the area of a minimal surface in the bulk, which is homogeneous to $A$. In the context of AdS$_3$/CFT$_2$ correspondence, the minimal surface in the bulk is just the geodesic connecting two endpoints of the spacial interval and the geodesic length gives the entanglement entropy of the interval in two dimensional(2D) CFT in the large central charge limit. This picture has been proved in \cite{Hartman:2013mia,Faulkner:2013yia}.

One interesting implication is from the holographic computation of the single interval entanglement entropy  for a 2D CFT on a circle at high temperature\cite{Takayanagi}. When the interval is short, the entropy could be read from the geodesics in the Banados-Teitelboim-Zanelli(BTZ) black hole background. However, when the interval is large, there could be two possibilities. One is the usual geodesic length, while the other one could be the sum of the BTZ black hole horizon length and the geodesic length of a very short interval complementary to the original one. In the large interval limit, the latter one dominates the contribution. This inspired the authors in  \cite{Takayanagi} to propose a universal relation between the thermal entropy and the entanglement entropy
\be\label{th} S_{th}=\lim_{l\rightarrow 0}(S_{EE}(R-l)-S_{EE}(l)). \ee
Actually, from a holographic computation, it was pointed out in \cite{Hubeny:2013gta} that if $l$ is just below a critical value, the Araki-Lieb inequality is saturated. However, this could only be  true in the large central charge limit.
The relation has been checked in the cases of a free fermion\cite{Takayanagi} and a noncompact free boson\cite{Datta}. It would be interesting to check this relation for general CFT.

In this letter, we prove the relation (\ref{th}) for the CFTs with a discrete spectrum.  In applying the replica trick to compute the entanglement entropy, we have to calculate the partition function of CFT on a higher genus Riemann surface coming from pasting the tori along the intervals. Here we present two proofs of the relation (\ref{th}), one  using the complete basis of normal sector states from multi-replica field theory, the other relying on  the complete basis from the twist sector states in orbifold CFT. In the latter case, the one-to-one correspondence between the twist sector states and normal sector states in an orbifold CFT  allows us to prove (\ref{th}). Moreover, we discuss the relation for a 2D noncompact free scalar. In this case, the theory is gapless and has a continuous spectrum.
The straightforward computation on the partition function suggests that there is a  log-logarithmic term in the large interval limit of the entanglement entropy. Such a term cannot be canceled and therefore leads to a mismatch. However, if we regularize the theory by taking it as the large volume limit of a compact scalar, and if we take the  limits appropriately, we recover the relation.

\section{First Proof }

To compute the entanglement entropy, it is convenient to use the so-called R\'enyi entropy, which is defined as
\be
S_A^{(n)}=-\f{1}{n-1} \log \Tr_A \r_A^n.
\ee
It gives the entanglement entropy $S_A=\displaystyle{\lim_{n \to 1} S_A^{(n)}}$,
if the analytic continuation $n \to 1$ limit is well-defined. By the replica trick\cite{Callan:1994py}, the R\'enyi entropy in two dimensional quantum field theory can be transformed into calculating the partition function on a higher genus Riemann surface,  \cite{Calabrese:2009qy}
\be S^n=-\frac{1}{n-1}\log\frac{Z_n}{Z_1^n}, \ee
where $Z_n$ is the partition function for $n$ tori connecting along the branch cut.    To calculate the partition function, we can cut the torus along a cycle and insert a complete basis there.

\begin{figure}[tbp]
\centering
\subfloat[]{\includegraphics[width=4cm]{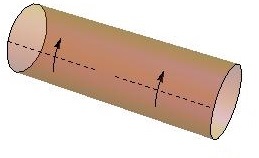}\label{i1}}
\quad
\subfloat[]{\includegraphics[width=4cm]{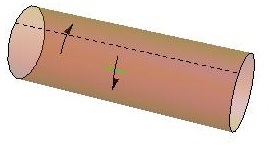}\label{i2}}
\\
\caption{Riemann surface for finite temperature large interval entanglement entropy. (a): The dashed line is the branch cut, and the arrow indicates the field  transforming from the $j$th to the $(j+1)$th replica.    (b): The modified branch cut. Note that one should identify the two ends of the cylinder to get a torus. }
\end{figure}

We consider a large interval on a circle of radius $R$. The interval length is $L=R-l$ with $l$ being very small $l/R <<1$. Without losing generality, we set the branch points at $u_1=\frac{l}{2}, u_2=-\frac{l}{2}$ such that the interval extends from $u_1$ to $u_2$ winding around the spacial cycle. In other words, the interval is the union  $[-\frac{R}{2},-\frac{l}{2}]\bigcup[\frac{l}{2},\frac{R}{2}]$. The CFT is at a finite temperature, with thermal radius being $\b=1/T$. We assume that the CFT has a discrete spectrum. As we are working with Euclideanized field theory, we are allowed to quantize the theory along the thermal direction or along the spacial direction. Let us first consider the quantization along the thermal direction. In this case, the thermal density matrix is of the form
\be
\r_{th}=e^{-\b H}
\ee
with the Hamiltonian 
\be
H=\frac{2\pi}{R}(L_0+\widetilde{L_0}-\frac{c}{12})
\ee
Then we have the thermal partition function
\be
Z_1=\Tr e^{-\b H}=e^{\frac{2\pi \b c}{12 R}}\sum_ie^{-\frac{2\pi \b}{R}\D_i}, \label{part1}
\ee
where the summation is over all the excited states with conformal dimension $\D_i$. To compute the partition function $Z_n$ on the $n$-sheeted Riemann surface, we should cut the cylinder along the spacial cycle and insert a set of complete bases of $n$-replica CFT\cite{Cardy2,small}. In this way, we find that
\bea
Z_n&=&\sum_{j_1,j_2,\cdots j_n}\langle j_1,j_2,\cdots j_n\mid {\cal{T}}^-(u_1){\cal{T}}^+(u_2)\mid j_1,j_2,\cdots j_n\rangle\mid_A\notag \\
&& \cdot \exp\left(-\frac{2\pi \b}{R}\sum_{j_i}\D_{j_i}+\frac{2\pi \b}{R}\frac{nc}{12}\right),\notag
\eea
where the summation is over all the excitations of CFT in every replica and $\D_{j_i}$ is the conformal dimension of the excitation in the $j_i$-th replica.
Here ${\cal{T}}^-(u_1)$ and ${\cal{T}}^+(u_2)$ are two twist operators inserted at the branch points of the interval. And the subscript $A$ means that the branch cut is along interval $A$ in \ref{i1}. Note that as we are working on a torus with a compact spatial direction, it is necessary to identify the branch cut. The above correlation function is defined on an $n$-sheeted cylinder pasted along the branch cut shown in Fig. \ref{i1}. As the interval is very large, these two twist operators are very close to each other. Actually, we may deform the branch cut in such a way that it extends the whole spatial cycle (dashed line in Fig. \ref{i2}), being subtracted by the complementary small interval (green line), as shown in Fig. \ref{i2}.  As a result,  we can rewrite $Z_n$ as
\bea
Z_n&=&\sum_{j_1,j_2,\cdots j_n}\langle j_1,j_2,\cdots j_n\mid {\cal{T}}^-(u_1){\cal{T}}^+(u_2)\mid j_2,\cdots j_n,j_1\rangle\mid_B \notag \\
&~&\cdot \exp\left(-\frac{2\pi \b}{R}\sum_{j_i}\D_{j_i}+\frac{2\pi \b}{R}\frac{nc}{12}\right),\notag
\eea
with $B=A^{c}$. Here the dashed line connects consecutive sheets, and transforms the states  into the next replica.  %Equivalently, the deformation of the interval could be understood as inserting two other twist operators at the end of the dashed line and at the same time exchange the twist operators at the original branch point. Consequently the two-point correlation functions changes to a four-point functions.
By the monodromy condition, we know that the twist operator should not change.
In the large interval limit, we may use the operator product expansion (OPE)
\be {\cal{T}}^{-}(-\frac{l}{2}){\cal{T}}^{+}(\frac{l}{2})\sim c_nl^{-\frac{c}{6}(n-\frac{1}{n})}(1+O(l)), \label{OPE}\ee
where we only keep the first term such that the correlation function of the twist operator reduces to the inner product of the complete basis, which is nonvanishing only when the excitations on different replica are the same. Consequently
\be
Z_n=e^{\frac{2\pi \b}{R}\frac{nc}{12}}c_nl^{-\frac{c}{6}(n-\frac{1}{n})}(\sum_ie^{-\frac{2\pi n\b}{R}\D_{i}}+O(l)),
\ee
from which we have
\bea S_n&=&-\frac{1}{n-1}\left(\log~(c_nl^{-\frac{c}{6}(n-\frac{1}{n})})+\log\frac{\sum_ie^{-\frac{2\pi n\b}{R}\D_{i}}}{(\sum_ie^{-\frac{2\pi \b}{R}\D_i})^n}\right) \notag \\
&~&+O(l).\notag
\eea
The quantity we are interested in is
\bea\label{EE1} \lefteqn{\lim_{l\rightarrow0}(S_{EE}(R-l)-S_{EE}(l))}\notag \\
&=&-\lim_{n\rightarrow1}\frac{1}{n-1}(\log~Z[\frac{n\beta}{R}]-n\log~Z[\frac{\beta}{R}]) \notag \\
&=&\log Z[\frac{\beta}{R}]-\frac{R}{\beta}\frac{Z^{'}[\frac{\beta}{R}]}{Z[\frac{\beta}{R}]},
\eea
with
\be Z[x]\equiv\sum_ie^{-2\pi x\Delta_i}. \ee
Note that the terms proportional to $c_n$ have been canceled.

On the other hand, the thermal entropy could be obtained by
\bea &~&S_{th}=-\frac{\partial F}{\partial T}=\log Z[\frac{\beta}{R}]-\frac{R}{\beta}\frac{Z^{'}[\frac{\beta}{R}]}{Z[\frac{\beta}{R}]},\label{th1}
\eea
which is the same as (\ref{EE1}). Therefore we prove the relation (\ref{th}).

We would like to emphasize that the above proof is valid for any temperature. From the holographic point of view, the absence of  the term linear in $c$ in the entanglement entropy seems to indicate that this discussion is only true in the low temperature limit. However, this is just an illusion. Our proof above is obviously independent of the temperature. One may worry that at high temperature one should quantize the theory along the spatial direction rather than the thermal direction. This worry is not necessary as we know that the partition function is modular invariant. Actually it is possible to discuss the problem from the quantization along the spacial direction, as we show below.

%If the interval is not very large,  we may cut the Riemann surface along the thermal cycle $A^{(i)}$, and insert all the states of normal sector of $n$ copies of CFT. Alternatively, we may cut the Riemann surface along the cycle crossing the branch cut. In the latter case, as the fields have nontrivial monodromy around the branch cut, we have to consider the twist sector in the orbifold CFT resulted from the replica trick.

\section{Second proof}

If we quantize the theory along the spacial direction, the density matrix is of the form
\be
\r_s=e^{-RH_s}=e^{-\frac{2\pi R}{\b}}(L_0+\widetilde{L_0}-\frac{c}{12}).
\ee
In this case, we need to cut the thermal cycle and insert a complete set of bases there.
However if we cut the thermal cycle across the branch cut, due to the presence of the interval, the field satisfies nontrivial monodromy condition. This requires us to insert a complete basis of twist sector states.

Suppose that one of the branch points is at the origin, then the field satisfies the twist boundary condition
\be\label{bc} \phi^{(j)}(ze^{2\pi i},\bar{z}e^{-2\pi i})=\phi^{(j+1)}(z,\bar{z}),  \ee
with $j=1,\cdots, n$  labeling the sheets, and $\phi$ can be any field in a CFT. We can redefine other $n$ fields as
\be \phi^{(t,k)}(z,\bar{z})=\sum_{j=1}^ne^{\frac{2\pi i}{n}kj}\phi^{(j)}(z,\bar{z}) \ee
with the monodromy condition
\be \phi^{(t,k)}(ze^{2\pi i},\bar{z}e^{-2\pi i})=\phi^{(t,k)}(z,\bar{z})e^{-\frac{2\pi ik}{n}}. \ee
The mode expansion of the field in the twist sector is
\be \phi^{(t,k)}=\frac{1}{n^{h+\bar{h}-1}}e^{\frac{2\pi i}{n}k}
\sum_{\substack{m=-h+s,\bar{m}=-\bar{h}+\bar{s} \\m-\bar{m}=k+an}}
\frac{\phi_m\bar{\phi}_{\bar{m}}}{z^{h+\frac{m}{n}}\bar{z}^{\bar{h}+\frac{\bar{m}}{n}}}\notag\ee The lowest state in the twist sector is
\bea \phi_{-h} \bar{\phi}_{-\bar{h}} \mid t \rangle,  \eea
with the conformal dimension $(\frac{h}{n}+\frac{c}{24}(1-\frac{1}{n^2}), \frac{\bar{h}}{n}+\frac{c}{24}(1-\frac{1}{n^2})$. Here $\mid t \rangle$ is the twist vacuum with conformal dimension $h=\frac{c}{24}n(1-\frac{1}{n^2})$, and the higher conformal dimension states can be built by acting creation operators on this state.
 There is a one-to-one correspondence between the twist sector states and normal sector states, with their energies being related by\cite{ChenWu2}
 \bea\label{en} H&=&\frac{2\pi}{\beta}(L_{twist}+\tilde{L}_{twist}-\frac{nc}{12})
=\frac{H_{normal}}{n}. \eea
With this correspondence, let us discuss the relation (\ref{th}) again.
The partition function $Z_1$ is now
\be
Z_1=e^{R\frac{2\pi}{\beta}\frac{c}{12}}\sum_ie^{-\frac{2\pi R}{\beta}\Delta_i}.
\ee
For the partition function on an $n$-sheeted Riemann surface, we have
\bea\notag Z_n&=&e^{\frac{\pi c}{6n}\frac{R}{\beta}}\sum_i\langle t,i\mid {\cal{T}}^-(-\frac{l}{2}){\cal{T}}^+(\frac{l}{2})\mid t,i\rangle
e^{-\frac{2\pi R}{n\beta}\Delta_i}, \eea
where the label $t$ denotes the twist sector, and the summation is over all the states in the twisted sector. As the branch points are near each other, we can still use the OPE of the twist operators (\ref{OPE}), from which we find
\bea S_n
&=&-\frac{1}{n-1}\left(\frac{\pi c}{6n}\frac{R}{\beta}+\log~(c_nl^{-\frac{c}{6}(n-\frac{1}{n})})
+\log(\sum_ie^{-\frac{2\pi R}{n\beta}\Delta_i})\right. \notag \\
&~&\left. -n(\frac{\pi c}{6}\frac{R}{\beta}+\log(\sum_ie^{-\frac{2\pi R}{\beta}\D_i})+O(l))\right).\notag
\eea
Consequently, we have
\bea\label{th2} \lefteqn{\lim_{l\rightarrow0}(S_{EE}(R-l)-S_{EE}(l))}\notag \\
&=&\frac{\pi c}{3}\frac{R}{\beta}+\log Z[\frac{R}{\beta}]+\frac{R}{\beta}\frac{Z^{'}[\frac{R}{\beta}]}{Z[\frac{R}{\beta}]}.
\eea
The thermal entropy is now
\bea &~&S_{th}=\frac{1}{3}\pi c\frac{R}{\beta}+\log Z[\frac{R}{\beta}]+\frac{R}{\beta}\frac{Z^{'}[\frac{R}{\beta}]}{Z[\frac{R}{\beta}]},
\eea
which is the same as (\ref{th2}). This completes our second proof of the relation (\ref{th}).

In the large central charge limit, the first term in (\ref{th2}) gives the entropy of the BTZ black hole in the bulk, which is dual to the CFT at high temperature. The remaining terms correspond to the quantum corrections.
However, we would like to emphasize again that the relation (\ref{th2}) is valid for all temperatures. It is equal to the relation (\ref{th1}), both of which give the thermal entropy of the CFT.

\section{A counter-example?}

In the above proof, we have assumed that the CFT has a discrete spectrum. It would be interesting to see if the relation (\ref{th}) is still true for the CFT with a continuous spectrum. Here we discuss the relation for a non-compact free scalar\cite{comment1}. The partition function of this case has been given in terms of $W$ functions in \cite{Datta}.  One may compute the large and short interval expansions of the partition function and check the relation directly, as discussed carefully in \cite{ChenWu2}. In the large interval expansion, there appears a term of the form $\log (|\log l|)$, which could not be canceled by the short interval terms. The reason for such a term can be found in the discussion above. One essential point in the above proofs is to use the OPE of the twist operators. However, in the case of a noncompact scalar, the OPE relation (\ref{OPE}) breaks down. Actually, the relation changes to
\bea\label{OPEc} &~& {\cal{T}}^+(-\frac{l}{2}){\cal{T}}^-(\frac{l}{2}) \notag \\
&=& c_nl^{-\frac{1}{6}n(1-\frac{1}{n^2})}\left\{V^{n-1}\int\prod_{j=1}^ndk_j\delta(\sum_{j=1}^n k_j)\right.\notag \\
&~&\cdot\prod_{j=1}^n(\frac{l}{n})^{2k_j^2}e^{-ik_j\phi^{(j)}(u)}\mid_{u=0} \notag \\
&~&\left.\cdot \prod_{1\leq j_1<j_2\leq n}(2\sin\frac{\pi}{n}(j_2-j_1))^{2k_{j_1}k_{j_2}}+O(l)\right\}.\notag
\eea
where V is the regularized volume in the target space. We regard the non-compact boson as a large volume limit of a compact boson; therefore,  we have
\be\label{regulation} \sum_k \rightarrow \frac{V}{2\pi}\int dk,~~~ \delta_{k_1,k_2} \rightarrow \frac{2\pi}{V}\delta(k_1-k_2),
~~~\delta(0)=\frac{V}{2\pi}. \ee
 Different from the small interval case, the operators $\displaystyle{\prod_{j=1}^ne^{ik_j\phi^{(j)}}}$  contribute to the partition function in the large interval limit. In the end, we find that
\be \lim_{l\rightarrow 0}S_{EE}(R-l)-S_{EE}(l)=\log(|\log l|)+S_{th}+c_1, \notag\ee
where $c_1=1+\log4\pi-2\log V$.
This result could be obtained by direct expansion of the $W$ functions in the partition function as well\cite{ChenWu2}. Actually in getting this result, we have always taken the $V\to \infty$ limit first and then $l\to 0$ such that the theory has a continuous spectrum. The log-logarithmic divergence stems from the continuous spectrum of the non-compact scalar. %It is remarkable that after removing this divergence, the relation (\ref{th}) still holds after absorbing the constant term by regularization.
%a continuous spectrum will contribute into the operator product expansion, and we will show more detail in \cite{ChenWu2}

In the above discussion, we treat the noncompact free scalar directly, which is gapless and has a continuous spectrum. However, if we start from a compact free scalar with radius $V$ and study the universal relation (\ref{th}), and finally take $V\to \infty$ limit, then the perplexing term $\log (|\log l|)$ does not appear. Actually, as shown in \cite{Chen:2015cna}, the universal relation (\ref{th}) indeed holds for the compact free scalar. Therefore, once we take the noncompact free scalar as a limiting case of compact free scalar, the universal relation still holds.

\section{Conclusion and discussion}

In this letter, we proved the universal relation (\ref{th}) between the thermal entropy and entanglement entropy for the CFTs with a discrete spectrum. We also discussed this relation for noncompact free scalar with a continuous spectrum and found that there was a subtle order-of-limits issue. If we regularize the theory and take limits in the correct order, the relation still holds. One interesting issue is to  check if the relation (\ref{th}) could be true for a generic 2D quantum field theory without conformal symmetry.

In higher dimensions, the recent study on the holographic entanglement entropy suggests that the Araki-Lieb inequality could be saturated, which is called the entropy plateau\cite{Hubeny:2013gta}. This suggests that there exists some kind of generalization of the relation (\ref{th}) in higher dimensions. It would be interesting to study such a relation directly in the field theory.

\vspace*{5mm}
\noindent {{\bf Acknowledgments}}
The work was  supported in part by NSFC Grant No.~11275010, No.~11335012, and No.~11325522.
\vspace*{3mm}

%\begin{appendix}
%\end{appendix}

%%%%%%%%%%%%%%%%% references %%%%%%%%%%%%%%%%%%%%%%%%%%%%%%%%%%%%%%%%%%%%%%%%%%%%%%%%%%%%%%%%%%%%%%%

%%%%%%%%%%%%%%%%%% references %%%%%%%%%%%%%%%%%%%%%%%%%%%%%%%%%%%%%%%%%%%%%%%%%%%%%%%%%%%%%%%%%%%%%%%%%%%%%%%%%

%\end{CJK*}

\begin{thebibliography}{99}
\bibitem{comment1} Our discussion is different from the one in \cite{Datta}, which based on an incorrect relation between $W$ functions in the partition function.
\end{thebibliography}
\end{document}